\documentclass{emulateapj}

\shorttitle{Semiregular variables}
\shortauthors{Soszy{\'n}ski \& Wood}

\begin{document}

\title{Semiregular variables with periods lying between \\
the period-luminosity sequences C$'$, C and D}

\author{I. Soszy{\'n}ski}
\affil{Warsaw University Observatory, Al.~Ujazdowskie~4, 00-478~Warszawa, Poland}
\email{soszynsk@astrouw.edu.pl}
\and
\author{P.~R. Wood}
\affil{Research School of Astronomy and Astrophysics, Australian National University, Cotter Road, Weston Creek ACT 2611, Australia}
\email{wood@mso.anu.edu.au}

\begin{abstract}
We analyze the distribution of semiregular variables and Mira stars in the
period--luminosity plane. Our sample consists of 6169 oxygen-rich
long-period variables in the Large Magellanic Cloud included in the
OGLE-III Catalog of Variable Stars. There are many stars with periods that
lie between the well known sequences C and C$'$. Most of these stars are
multi-periodic and the period ratios suggest that these stars oscillate in
the same mode as the sequence C stars. Models suggest that this mode is the
fundamental radial pulsation mode. The stars with primary periods between
sequences C and C' preferentially lie on an additional sequence (named F),
and a large fraction of these stars also have long secondary periods that
lie between sequences C and D. There are also a small number of stars with
primary periods lying between sequences C and D. The origin of this long
period variability is unknown, as is the cause of sequence D
variability. In addition, the origin of sequence F is unknown but we
speculate that sequence F variability may be excited by the same phenomenon
that causes the long secondary periods.
\end{abstract}


\section{Introduction}

Variability of red giant stars is one of the least understood aspects
of stellar astrophysics. It is known that all stars in the upper part
of the first ascent red giant branch (RGB) and on the asymptotic giant
branch (AGB) vary, and the amplitude of variation tends to increase
toward higher bolometric luminosity. Traditionally, long-period
variables (LPVs) have been divided into three types: Miras,
semiregular variables (SRVs) and irregular variables \citep[see the
General Catalogue of Variable Stars, GCVS;][]{kholopov1985}. The
existence of irregular variables (i.e. variable stars with no sign of
periodicity) among red giants is in dispute
\citep{kerschbaum2001,lebzelter2009}. On the other hand, a very
numerous group of pulsating red giants, called OGLE Small Amplitude
Red Giants \citep[OSARGs;][]{wray2004}, seem to constitute a separate
class of variable red giant, different from Miras and SRVs
\citep{soszynski2004a}.

\begin{figure*}
\epsscale{0.80}
\plotone{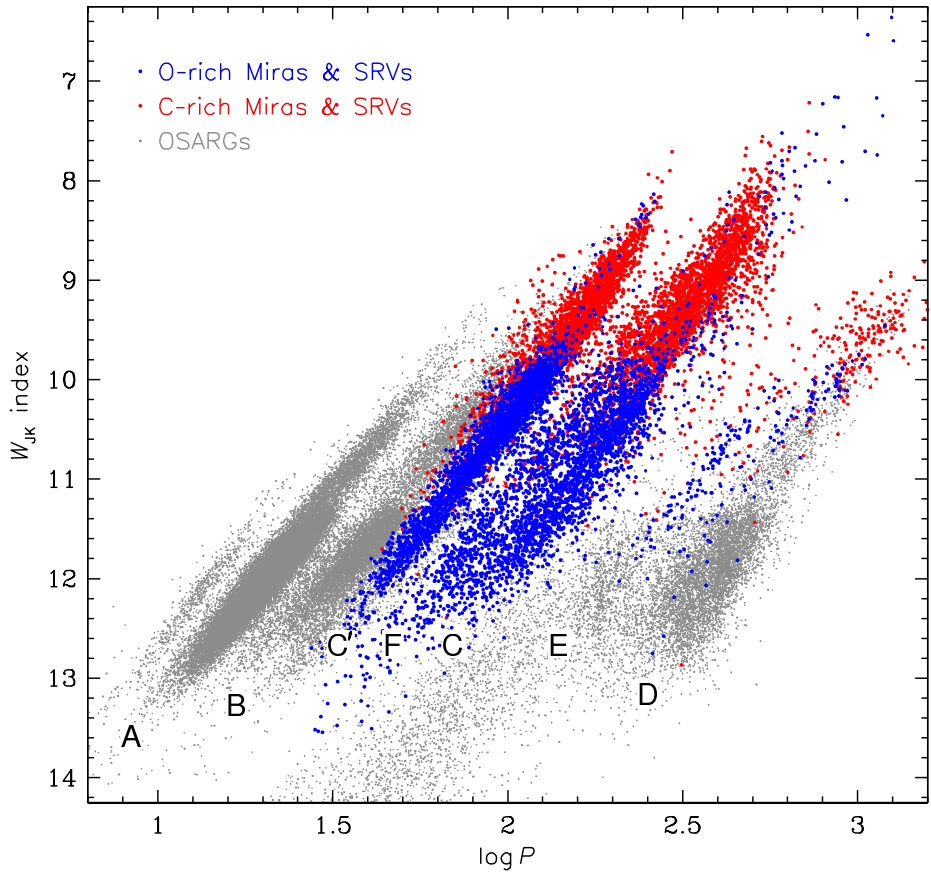}
\caption{Period--luminosity diagram for LPVs in the LMC. Each star is
represented by one point, corresponding to the primary (dominant)
period. Different colors refer to different types of variable sources:
blue points are O-rich SRVs and Miras, red points indicate C-rich SRVs
and Miras, and grey points show OSARG variables. Periods used for the
sequence E stars (binary systems) are half of the orbital periods.
Periods and classification of stars were taken from the OGLE-III Catalog
of Variable Stars \citep{soszynski2009}, while $J$ and $K$ magnitudes
used to construct the Wesenheit index originate from the 2MASS All-Sky
Point Source Catalogue \citep{cutri2003}.
\label{fig1}}
\end{figure*}

In the classical picture, semiregular red giants are sub-classified
into SRa and SRb stars. The definitions provided in the GCVS do not
give strict criteria to distinguish between both subtypes. SRa stars
display persistent periodicities with variable amplitudes and light
curve shapes, while SRb stars have poorly defined periodicities or
show alternating intervals of periodic and slow irregular changes. So
far, it has not been clear whether both types form a continuum, or
represent distinct classes of LPVs.

In recent years our knowledge of the distribution of LPVs in the
period--luminosity (PL) plane has significantly expanded. Before the
era of large microlensing surveys, two PL relations obeyed by LPVs 
at near-infrared (NIR) wavelengths were known: one discovered by
\citet{glass1981} for Mira stars and the second noticed by
\citet{wood1996} for SRVs. The long-term, massive photometry obtained
by the microlensing projects (MACHO, OGLE, EROS) together with
observations from the large-scale NIR surveys (e.g. 2MASS, DENIS,
IRSF) showed a very complex pattern formed by LPVs in the PL
plane. \citet{wood1999} and \citet{wood2000} distinguished and labeled
with letters A--E five PL ridges. Later studies
\citep{kiss2003,ita2004,soszynski2004a,soszynski2005,soszynski2007,fraser2008}
increased to fourteen the number of known PL relations obeyed by LPVs
when separate sequences are assigned to C stars and O-rich stars, and
to AGB and RGB stars \citep[see Table~1 in][]{soszynski2007}.
Furthermore, some of these sequences, especially sequences A and B, 
appear to be split into two or three close parallel sequences.
 
Fig.~\ref{fig1} presents the PL diagram for LPVs in the Large Magellanic
Cloud (LMC). The data originate from the OGLE-III Catalog of Variable
Stars\footnote{The catalog data are publicly available from the FTP site
ftp://ftp.astrouw.edu.pl/ogle/ogle3/OIII-CVS/lmc/lpv/ or via the web page
http://ogledb.astrouw.edu.pl/{\textasciitilde}ogle/CVS/}
\citep{soszynski2009}. Using the notation introduced by
\citet{wood1999} and other authors, Miras belong to sequence C, while
SRVs occupy sequences C and C$'$. OSARG variables populate sequences
A, B and several additional ridges \citep{soszynski2004a}. The OSARG
variables on the RGB and on the AGB follow somewhat different sets of
PL relations, shifted in period relative to each other
\citep{kiss2003}. Sequence D corresponds to the long secondary periods
(LSPs) -- a still unexplained phenomenon existing in at least one
third of SRVs and OSARGs
\citep[e.g.][]{wood2004,nicholls2009}. Sequence E consists of
eclipsing or ellipsoidal close binary systems with a red giant as one
of the components \citep{wood1999,soszynski2004b,nicholls2010}.

\citet{soszynski2005} noticed an additional PL ridge spreading between
sequences C and C$'$. This new PL relation is obeyed by SRVs with
relatively small amplitudes. The existence of this additional sequence
was confirmed for rich samples of LPVs in the LMC
\citep{soszynski2009} and SMC \citep{soszynski2011}. Here we
analyze in detail red giants lying on this sequence and generally
between sequences C and C$'$.

\section{Observational Data}

In this work, we restrict our analysis to the sample of LPVs in the LMC,
since this galaxy hosts very large numbers of AGB stars with well defined
multiple PL relations. The collection of LPVs was extracted from the
OGLE-III Catalog of Variable Stars \citep[OIII-CVS;][]{soszynski2009}. The
light curves published in the catalog were obtained with the 1.3-m Warsaw
telescope at Las Campanas Observatory, Chile, over the period June 2001 --
May 2009 (the time-span of the OGLE-III survey). The telescope and the
instrumentation are described in detail by \citet{udalski2003}. For
stars in the central part of the LMC, the OGLE-II observations
obtained from January 1997 to November 2000 were also included. About
90\% of the OGLE photometric measurements were collected in the
Cousins {\it I}-band, and we used these data to analyze periodicities
and amplitudes of our sample. Here we concentrate on the oxygen-rich
(O-rich) SRVs and Miras. Note that we have adopted the definition of
SRVs and Miras given by \citet{soszynski2007}: they have periods
located on sequences C and/or C$'$ but do not have periods on any of
the shorter period sequences (stars with periods on any of the shorter
period sequences are designated as OSARG variables). We decided not to
include carbon-rich (C-rich) LPVs, because they occupy the upper part
of the PL diagram (Fig.~\ref{fig1}) and sequence F is most easily
distinguishable for the fainter O-rich stars.

\begin{figure*}
\epsscale{0.90}
\plotone{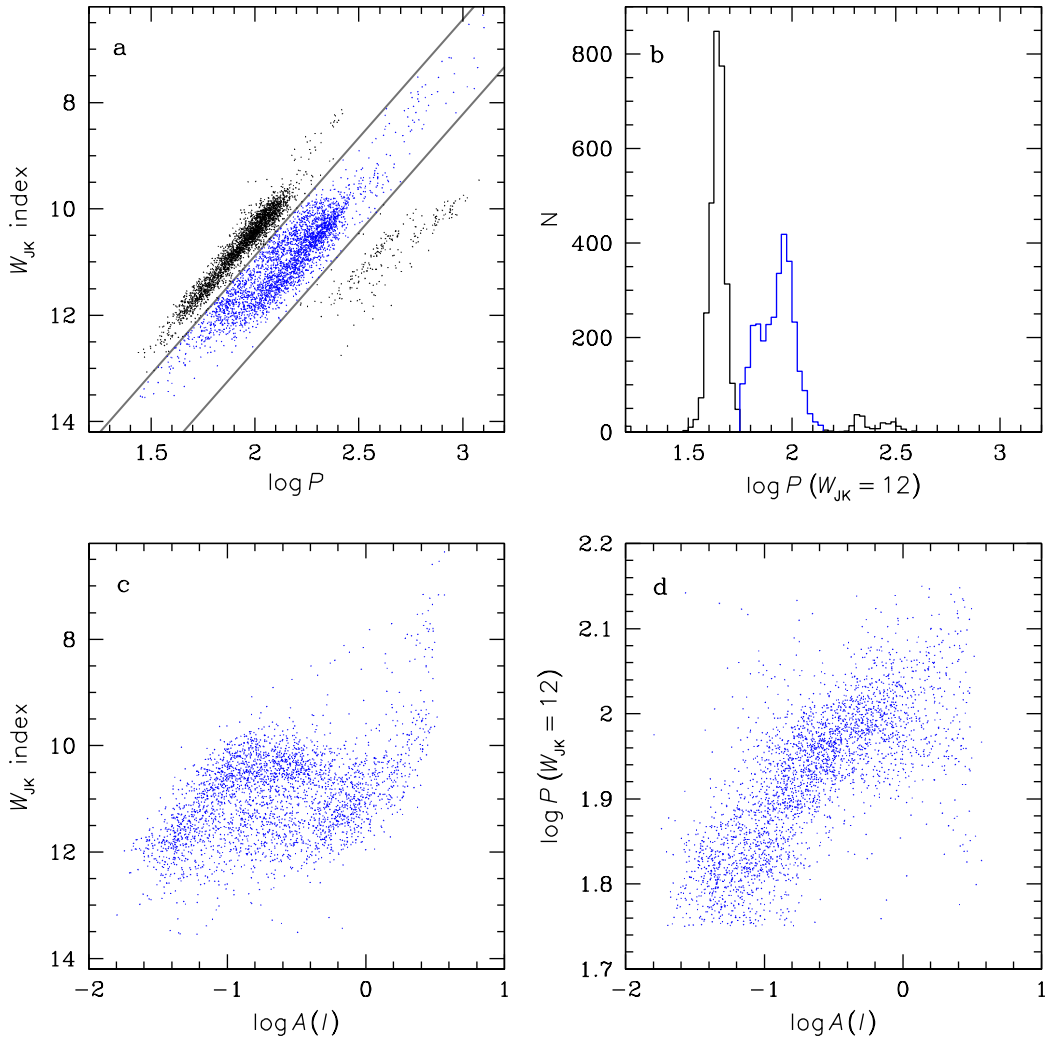}
\caption{Panel $a$: The $W_{\rm JK}$-$\log P$ diagram for the O-rich SRVs
and Miras in this study, where $P$ is the primary period of variation. The
two diagonal lines form the boundary of the region occupied by stars
between sequences C$'$ and C. Panel $b$: The histogram of the number of
stars as a function of $\log P(W_{\rm JK}=12)$, where $\log P(W_{\rm
JK}=12)$ is the value of $\log P$ obtained by projecting a star's true
$\log P$ to $W_{\rm JK} = 12$ along a line parallel to those shown in panel
$a$. Panel $c$: The $W_{\rm JK}$-$\log A(I)$ plot for the stars between the
two diagonal lines in panel $a$, where $A(I)$ is the full I amplitude of
the primary period of variation. Panel $d$: The $\log P(W_{\rm
JK}=12)$-$\log A(I)$ diagram for these stars.
\label{fig2}}
\end{figure*}

The OGLE catalog of LPVs in the LMC contains in total 6445 O-rich SRVs
and Miras. In the present analysis we re-derived periods provided in
the catalog, with the iterative procedure of fitting and subtracting
the irregular slow variations. We used the {\sc Fnpeaks} code by
Z.~Ko{\l}aczkowski, which is based on Fourier analysis. For each
object we found the three most significant periods. All light curves
folded with the primary (largest amplitude) periods were visually
inspected and, in some cases, the periods were judged to be spurious
and were replaced by one of the additional periodicities. The
additional periods were not inspected and in some individual cases
they may be spurious, although the statistical properties of the
sample should still be visible with these automatically derived
periods. The peak-to-peak amplitudes $A(I)$ were derived by fitting
a third order Fourier series to the detrended {\it I}-band light
curves. Following \citet{soszynski2009}, we define Mira stars as those
having {\it I}-band amplitudes larger than 0.8~mag, although we did
not notice any clear natural boundary between O-rich Miras and SRVs,
as found for C-rich LPVs by \citet{soszynski2009}. We were left with
5712 SRVs and 457 Miras (96\% of the initial sample) with relatively
well defined primary periods. For over 90\% of stars in our sample we
obtained practically the same (differing by less than 2\%) primary
periods as provided in the OIII-CVS.

NIR $J$ and $K_s$ measurements of the stars were obtained from the
2MASS All-Sky Point Source Catalogue \citep{cutri2003}. To avoid the
effects of interstellar and circumstellar extinction we constructed
the reddening-free Wesenheit index, defined as:
$$W_{\rm JK}=K_s-0.686(J-K_s)$$
where the coefficient $0.686=A_K/E(J-K)$ was derived assuming the
extinction law provided by \citet{schlegel1998}. The distribution of LPVs
in the period--$W_{\rm JK}$ plane is very similar to the distribution in
the period--$K_s$ plane, with the exception of LPVs with heavy
circumstellar extinction \citep[usually C-rich Miras, e.g.][]{ita2011},
which show significantly lower brightness in the $K_s$ band than expected
from a linear extension of the period--$K_s$ relation.

\begin{figure*}
\epsscale{0.80}
\plotone{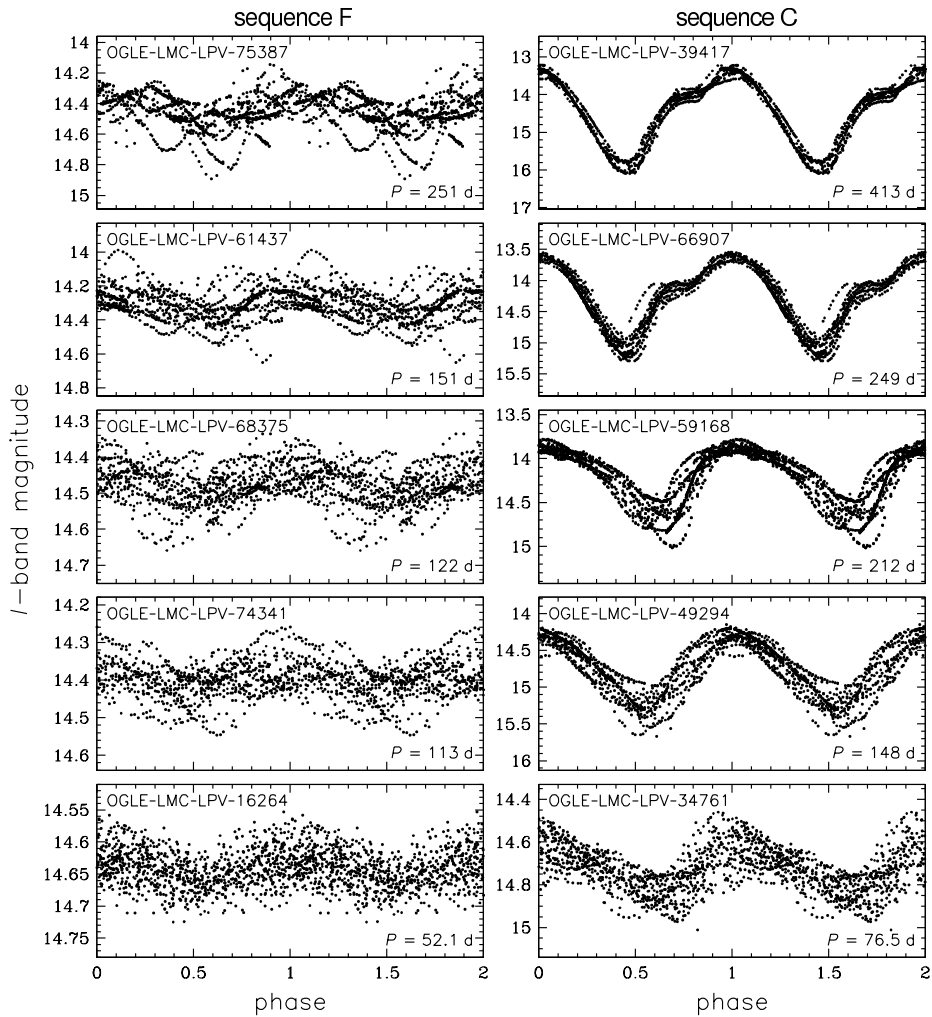}
\caption{Example {\it I}-band light curves of SRVs with the primary periods
lying on PL sequence F (left panels) and SRV/Mira stars which populate
sequence C (right panels). The light curves are folded with the periods
given in each panel.}
\label{fig3}
\end{figure*}

\section{Results and Discussion}

\subsection{The amplitudes and shapes of the light curves}
The goal of our study is to learn more about SRVs lying between sequences C
and C$'$. The sample of stars we examined was delimited in the
$\log{P}$--$W_{\rm JK}$ diagram by the two lines of slope d$W_{\rm
JK}$/d$\log{P}$ = 4.444 which are shown in panel $a$ of
Fig.~\ref{fig2}. Panel $b$ of Fig.~\ref{fig2} shows the histogram of the
number of stars horizontally across the strip between the two lines in
panel $a$. We use the parameter $\log P(W_{\rm JK}=12)$ to measure the
horizontal position of a point, where $\log P(W_{\rm JK}=12)$ is the value
of $\log P$ obtained by projecting a star's true $\log P$ down or up to
$W_{\rm JK} = 12$ along a line parallel to those shown in panel $a$ of
Fig.~\ref{fig2}. There are two peaks in this histogram, one with $\log
P(W_{\rm JK}=12) \approx 1.95$ which corresponding to sequences C and a
second weaker peak at $\log P(W_{\rm JK}=12) \approx 1.82$. Hereafter, we
refer to this ridge as sequence F, since F is the next letter (in
alphabetical order) after the series of designations introduced by
\citet{wood1999}\footnote{\citet{ita2004} used label F to designate the PL
relation for classical Cepheids, but we propose to continue this naming
scheme only for red giant stars}.

The amplitudes of our sample of stars are plotted against $W_{\rm JK}$ and
$\log P(W_{\rm JK}=12)$ in panels $c$ and $d$ of Fig.~\ref{fig2},
respectively. Panel $d$ shows that the amplitude of variation increases
steadily with $\log P(W_{\rm JK}=12)$, corresponding to moving from
sequence F to sequence C. The two sequences F and C can be clearly seen in
panel $c$ where the sequence F stars lie near the low amplitude edge of the
populated region while the sequence C stars lie near the high amplitude
edge. At brighter luminosities ($W_{\rm JK} \sim 10.5$) the amplitudes of
the sequence F stars appear to increase towards the amplitudes typical of
sequence C stars. These stars are possibly at the stage of increasing their
surface carbon abundance and will soon evolve into carbon-rich variables of
sequence C (the carbon stars merge with the O-rich stars at $W_{\rm JK}
\sim 10$).

Although the exact correspondence between the stars studied here and the
classical designations SRa and SRb is unclear, the smaller amplitudes of
the sequence F stars suggests that they may belong to the SRb class while
the sequence C stars belong to the SRa class. Several example light curves
of stars from the two sequences are shown in Fig.~\ref{fig3}. The
larger-amplitude sequence C variables are Miras or SRVs with well-defined
periods, so they can readily be classified as Mira and SRa stars. The
lower-amplitude sequence F stars generally have much more poorly expressed
periodicity, with strong irregular components in the light curves,
characteristics typical of SRb stars.

\subsection{The nature of the variable stars between sequences C$'$ and C}

\begin{figure}
\epsscale{1.22}
\hspace{-4mm}\plotone{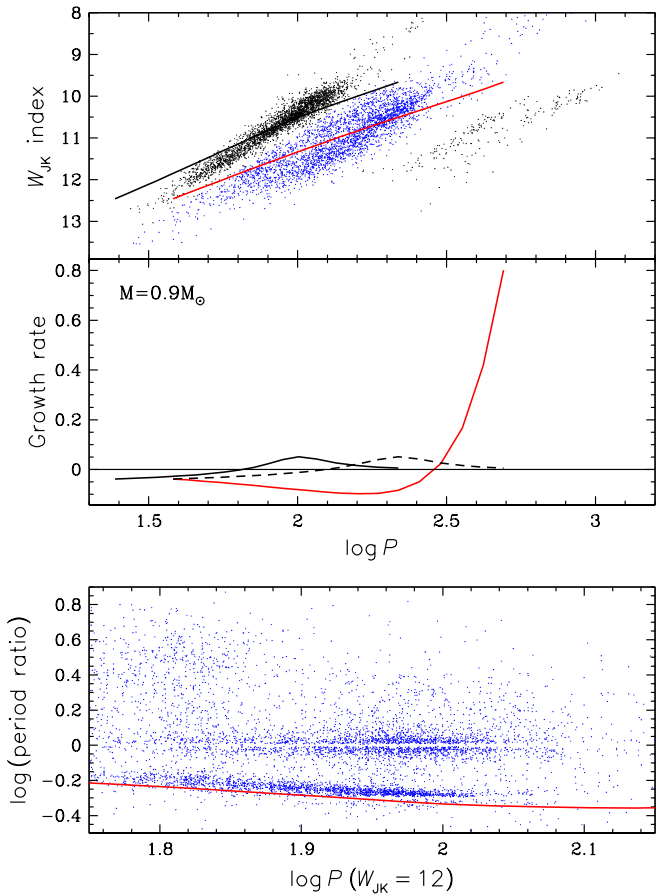}
\caption{Top panel: the $W_{\rm JK}$-$\log P$ diagram on which are
superimposed the positions of 0.9 M$_{\odot}$ radial pulsation models for
the fundamental mode (red) and the first overtone (black). Middle panel:
The growth rates (fractional increase in amplitude per period) for the
fundamental radial pulsation mode (red) and the first overtone radial
pulsation mode (black). The solid lines show the growth rate plotted
against the mode period while the dashed line shows the first overtone
growth rate plotted against the fundamental mode period (thus points on the
red line and the dashed black line with the same period correspond to the
same luminosity and model). Bottom panel: the period ratios $P_{\rm
2}/P_{\rm 1}$ and $P_{\rm 3}/P_{\rm 1}$ plotted against $\log P_{\rm
1}(W_{\rm JK}=12)$, where $P_{\rm 1}$ is the primary observed period of
variation and $P_{\rm 2}$ and $P_{\rm 3}$ are the secondary and tertiary
observed periods. Also shown is the ratio of the first overtone period to
the fundamental mode period for the 0.9\,M$_{\odot}$ models plotted against
$\log P(W_{\rm JK}=12)$ for the fundamental mode (red line).}
\label{fig4}
\end{figure}

In order to understand the nature of the variable stars between sequences
C$'$ and C, we compare observations with models in Fig.~\ref{fig4}. In
particular, we examine the belief that the cause of variability in these
stars is predominantly due to pulsation in radial modes\footnote{We note that CoRoT \citep[e.g.][]{mosser2010} and Kepler \citep[e.g.][]{kallinger2010} observations of red giants about one order of magnitude fainter than our sample are yielding data on many convectively-excited radial and non-radial modes. At this stage, differential frequency or period information from the space-based observations are being used to derive information about the global and interior properties of the stars. The actual radial order of the modes is not precisely known. The present study of the most luminous giants aims to model the values of the periods of known radial modes in terms of global stellar properties. As the space-based observation sequences become longer and the luminosities of the red giants with detected frequencies begin to overlap with the giants observed by OGLE, we can expect a merger of these two approaches.}
\citep{wood1996,wood1999,soszynski2007}. The linear, non-adiabatic, radial
pulsation models used here were constructed with the code described by
\citet{fox1982} and updated by \citet{keller2006}. An additional update for
the current models is the use of opacities from \citet{marigo2009} which
include a molecular component in the outer layers. The models use a
composition Y=0.73 and Z=0.008, a mixing length of 1.97 pressure scale
heights (to reproduce the giant branch temperature given by
Kamath et al. 2010 for the luminous O-rich stars in the populous
intermediate age cluster NGC\,1978) and a turbulent viscosity parameter
$\alpha_{\rm \nu}$ = 0.25.

The top panel of Fig.~\ref{fig4} shows the observed stars in the
$W_{\rm JK}$-$\log P$ diagram. Also shown are the positions occupied by
0.9~M$_{\odot}$ stars pulsating in the fundamental and first overtone
modes. The models have been adjusted to the LMC using the distance modulus
18.54~mag and reddening $E(B-V)=0.08$~mag given by \citet{keller2006}.

The first thing to notice about this diagram is that the lines
corresponding to a fixed mode and a fixed mass are not parallel to the
observed sequences. A given star will evolve along these lines (neglecting
mass loss, which should be a minor effect for these optically visible
stars). The observed variability sequences are defined by the positions on
these lines at which the mode is unstable, and by the range of mass values
in the stellar population being considered (higher mass stars have tracks
to the left of, or equivalently above, the lines shown, and vice versa for
lower mass stars).

The computed stability of the modes in red giant stars is very uncertain
due to the fact that convection carries most of the flux through most of
the envelope and our knowledge of the transport of energy by convection is
poor (the mixing-length theory of convection is used here). Nevertheless,
the computed growth rates (middle panel of Fig.~\ref{fig4})
qualitatively confirm the above picture. It can be seen that the peak in
the growth rate for the first overtone mode occurs at the periods where the
black line crosses sequence C$'$, as required. The fundamental mode only
becomes unstable at the highest luminosities, again consistent with
observations. However, in the models the fundamental mode becomes unstable
at the long period edge of sequence C rather than on the short period
edge. This is clearly a quantitative deficiency in the models. Note that
the right hand edge of sequence C is defined not by the fundamental mode
becoming stable again but by loss of the stellar envelope caused by large
amplitude Mira-like pulsation. This general pattern of modal instability
and mass loss has been described before by \citet{wood1983} and
\citet{vassiliadis1993}.

Using the above picture, the evolution of AGB stars can be described
qualitatively as follows. A 0.9~M$_{\odot}$ star evolving up the AGB will
(neglecting helium shell flashes) move to higher luminosities until its
first overtone mode becomes unstable at which time it will belong to the
left side of sequence C$'$. With further evolution, the star will move to
higher luminosities and through to the right side of sequence C$'$. The
growth rates shown in Fig.~\ref{fig4} suggest that the growth rate, and
presumably the amplitude of pulsation, will be less there than when the
star was near the centre of sequence C$'$. Eventually, the star will
become unstable in the fundamental mode and the fundamental mode will
become dominant. At this stage, the star will make a transition to sequence
C at constant luminosity.

Exactly, how this transition occurs is quite uncertain. The growth rates
shown in Fig.~\ref{fig4} suggest that the first overtone remains
unstable when the fundamental mode becomes unstable suggesting that both
modes will be active simultaneously at certain luminosities (the red and
dashed curves in Fig.~\ref{fig4} should be used to compare the growth
rates in the same star). However, the computed growth rates are very
uncertain and, in particular, the (totally unknown) values of the turbulent
viscosity parameter $\alpha_{\rm \nu}$ can be used to suppress the growth
rates by varying amounts. It is even possible that after the star has
evolved through the region where the first overtone is unstable, both this
mode and the fundamental mode will be stable over a small luminosity
interval. In this case, any modes that are seen would be excited
externally, for example by convective motions or some other mechanism. This
possibility is discussed more in the next sub-section.

The bottom panel of Fig.~\ref{fig4} plots the period ratios $P_{\rm
2}/P_{\rm 1}$ and $P_{\rm 3}/P_{\rm 1}$ against $\log{P_{\rm 1}}(W_{\rm JK}=12)$ 
(we remind the reader that up to 3 periods were determined for each star,
with the primary period $P_{\rm 1}$ corresponding to the largest amplitude
period). Note that the range of $\log{P_{\rm 1}}(W_{\rm JK}=12)$ shown
corresponds to the complete width of the region containing sequences F and
C (the blue points in the top panel). We will hereafter designate stars
with $1.75 < \log{P_{\rm 1}}(W_{\rm JK}=12) < 1.9$ as sequence F stars and stars
with $1.9 < \log{P_{\rm 1}}(W_{\rm JK}=12) < 2.15$ as sequence C stars. The empty
strip in the bottom panel of Fig.~\ref{fig4} at $\log(period ratio) = 0.0$
is the zone where additional modes are unresolvable for the current length
of the OGLE data series. Similar empty strips occurs at period ratios of
1/2 and 1/3 since each period is fitted by the primary frequency and two
harmonics. Points with $-0.1 \la \log(P_{\rm 2}/P_{\rm 1}) \la 0.1$ or
$-0.1 \la \log(P_{\rm 3}/P_{\rm 1}) \la 0.1$ presumably correspond to
multiple detections of the same mode. It is well known that the periods of
long period variables vary erratically (e.g. see Fig. 10 of
Wood et al. 2004 or Templeton et al. 2005) so that the same mode may be
detected by Fourier analysis multiple times due to period variability.

There is a well defined locus of points on a line with $-0.3 \la
\log(P/P_{\rm 1}) \la -0.15$ corresponding to the simultaneous existence of
two modes. The red line in Fig.~\ref{fig4} shows the ratio of the first
overtone period to the fundamental mode period. This line follows the
general shape of the observed locus of points (although offset by a small
amount indicating that the structure of red giant models constructed using
the mixing-length theory of convection and extant opacities is not a good 
match to the structure of real stars). It
seems clear from the near-coincidence of the models and observations that
all the points on the observed locus correspond to stars in which both the
fundamental and first overtone radial modes are active. The lack of a
corresponding locus of points at positive values of $\log(P/P_{\rm 1})$
shows that the fundamental mode oscillation dominates the first overtone
oscillation for sequence F stars as well as for sequence C stars
i.e. $P_{\rm 1}$ in these stars is the fundamental radial mode. The
fraction of sequence C stars with a detected first overtone pulsation is
49\% and the fraction of sequence F stars with a detected first overtone
pulsation is 56\%. It is therefore reasonable to assume that all the stars
on sequences F and C are fundamental mode pulsators.

The above analysis does not give any reason for expecting a second sequence
for fundamental mode pulsation i.e. sequence F in addition to sequence C,
as suggested by the histogram in panel $b$ of Fig.~\ref{fig2}. One possible
way to produce sequence F would be to have a positive bump in the growth
rate of the fundamental mode near $\log P \sim 2$ in the middle panel of
Fig.~\ref{fig4}. However, in our models, the growth rate behaves
smoothly with $\log P$ so the present models can not explain sequence F in
this way.

\begin{figure}
\includegraphics[width=0.48\textwidth]{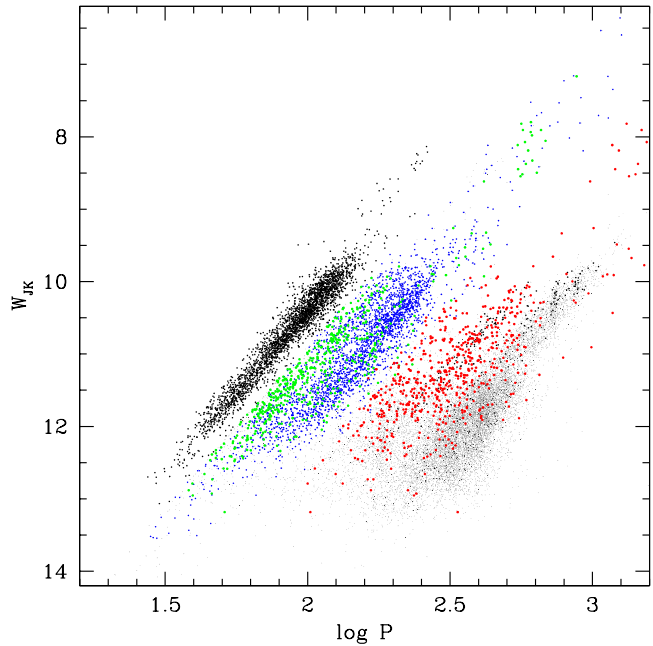}
\caption{The $W_{\rm JK}$-$\log P$ diagram for the O-rich SRVs and Miras in
this study, where $P$ is the primary period of variation. Blue points are
stars with their primary pulsation periods on sequences F or C. Green
points are sequence F or C stars that also have an additional period longer
than sequence C, where these additional periods are shown as red
points. The black points are stars whose primary periods lie on sequence
C$'$ or stars with primary periods longer than the sequence C periods. The
faint grey points are all the OSARG variables from the OGLE III catalog of
LPVs that have a primary period longer than sequence C. They are plotted to
show the position of sequence D.}
\label{fig5}
\end{figure}

\subsection{Long secondary periods in sequence F stars}

\begin{figure}
\epsscale{1.00}
\plotone{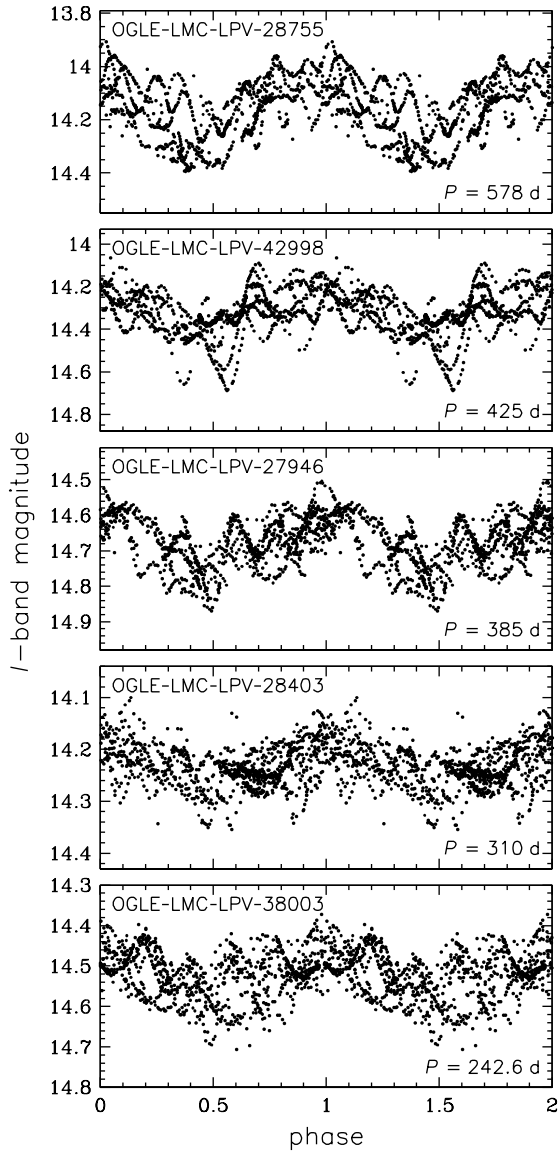}
\caption{Example {\it I}-band light curves of stars with the primary
periods falling between sequences C and D. The light curves are folded
with the periods given in the panels.}
\label{fig6}
\end{figure}

A notable feature in the bottom panel of Fig.~\ref{fig4} is the
considerable fraction (33\%) of sequence F stars in which $P_2$ or
$P_3$ is a long secondary period (LSP) with $\log(P/P_{\rm 1}) > 0.3$.
A much smaller fraction (8\%) of sequence C stars have LSPs defined in
this way.

Fig.~\ref{fig5} shows the positions of the LSPs associated with
sequence F and C stars as red points in the $W_{\rm JK}$-$\log P$
diagram. The stars on sequences F or C that host these LSPs are
marked as green points. It is clear from Fig.~\ref{fig5} that
the LSPs lie predominantly between sequence C and sequence D. Note
that there is also a small number of variables that have their primary
periods in this region (see Fig.~\ref{fig1}, Fig.~\ref{fig5} or
panel $a$ of Fig.~\ref{fig2}). These stars seem to fall on a quite
tight sequence and some examples of their light curves are shown in
Fig.~\ref{fig6}. We do not know the origin of either the primary
periods or the LSPs which lie between sequences C and D \citep[nor do
we understand the origin of the LSPs associated with sequence D -
see][] {wood2004,nicholls2009}. The concentration of green points
on sequence F suggests that there is some relation between the
sequence F fundamental mode oscillations and the LSPs lying between
sequences C and D. Perhaps the LSPs excite the fundamental mode
pulsation seen on sequence F.

It is very notable that the ratio of the LSP to the fundamental mode
period is not precisely defined, in contrast to the ratio of first
overtone period to the fundamental mode period (see the bottom panel
of Fig.~\ref{fig4}). This suggests that the long secondary periods
are not caused by one (or a few) pulsation modes. Some completely
separate mechanism must be involved.

In their study of 247 red giants in the solar neighborhood,
\citet{tabur2010} reported the discovery of LPVs with periods falling
between sequences C and D. They attributed these periodicities to the
stellar noise from convection. However, this is not the only possible
explanation of the physical origin of these periods. Another possible
explanation assumes that these are binary systems, since these objects
are on the extension of sequence E populated by ellipsoidal and
eclipsing binaries. In such a case their orbital periods should be two
times longer than the photometric period shown in the PL diagram
\citep{soszynski2004b,nicholls2010}. However, their light curves do
not exhibit characteristic features of ellipsoidal or eclipsing
variables (alternating deep and shallow minima and/or eclipses).
Fig.~6 shows an example set of the light curves folded with the
photometric periods falling between sequences C and D. Although
ellipsoidal modulation is not excluded, there is no clear evidence for
binarity as seen in the case of the sequence E stars. Other possible
non-pulsating explanation for the periods that lie between sequences C
and D are rotation of a spotted star and star spot cycles
\citep{wood2004} and a long period convective cycle
\citep{stothers2010}. It is possible that the same physical mechanism
applies to all the stars with periods longer than sequence C, whether
these periods lie on sequence D or between sequences C and D.

\section{Conclusions}

We have examined the distribution of semiregular variables and Mira
variables in the $W_{\rm JK}$-$\log P$ diagram. These stars all have
periods on, or longer than, those of sequence C$'$. Although the
majority of variables have periods on the well known sequences C and
C$'$, many variables lie between sequences C and C$'$ and a small
number have primary periods between sequences C and D. There appears
to be a concentration of stars on a faint sequence roughly half way
between sequences C and C$'$: we label this sequence F. These stars
have small amplitudes, approximately 0.1 times those of stars on
sequence C, and poorly defined periodicity so they would be defined as
SRb variables in the classical subdivisions of LPVs given in the
General Catalog of Variable Stars. For variables between sequences
C$'$ and C, there is a general increase in amplitude with luminosity
and with increasing period while at constant luminosity. Many of the
stars with primary periods on sequence F have secondary periods lying
between sequences C and D.

Most of the stars with primary periods lying between sequences C and
C$'$ are multimode variables. Comparison of the period ratios in
these stars with theoretical pulsation models shows that the primary
period is the period of the fundamental mode of radial pulsation.
Approximately 50\% of the stars have detected pulsations in the first
overtone mode. The growth rates in the models qualitatively predict
the existence of sequences C and C$'$, these being the regions of
instability of the fundamental and first overtone modes, respectively.
However, there is no evidence in the models for an increase in the
growth rate in the vicinity of sequence F. We are thus unable to
explain the origin of this sequence as due to self-excited pulsation.
We speculate that these stars may have their fundamental mode excited
by the phenomenon that generates the long secondary periods seen
between sequences C and D. The origin of these LSPs, and the LSPs of
sequence D, is unknown.

\section*{Acknowledgments}

We would like to thank W.~A.~Dziembowski for helpful comments.
The research leading to these results has received funding from the
European Research Council under the European Community's Seventh Framework
Programme (FP7/2007-2013)/ERC grant agreement no. 246678. This work has been
supported by the Polish National Science Centre grant
no. DEC-2011/03/B/ST9/02573. PRW acknowledges partial support provided by
Australian Research Council Discovery Project DP120103337.
He is also grateful for the hospitality provided by the Department of
Astronomy, University of Padova where much of his work on this paper
was done.

\label{lastpage}

\end{document}